\documentstyle[preprint,aps,epsf]{revtex}

\begin{document}

\newcommand{\be}{\begin{equation}}
\newcommand{\ee}{\end{equation}}
\newcommand{\bea}{\begin{eqnarray}}
\newcommand{\eea}{\end{eqnarray}}

\title {KMS conditions for 4-point Green  functions at finite temperature}

\author{M. E. Carrington${}^{a,b}$, Hou Defu${}^{a,b,c}$ and J. C. Sowiak${}^{a,d}$}
\address{
${}^a$ Department of Physics, Brandon University, 
Brandon, Manitoba, R7A 6A9 
Canada\\
${}^b$ Winnipeg Institute for Theoretical Physics, 
Winnipeg, Manitoba, R3B 2E9 Canada\\
 ${}^c$ Institute of Particle Physics,
Huazhong Normal University, 430070 Wuhan, China\\
${}^d$ Department of Mathematics, Brandon University, Brandon, 
Manitoba R7A 6A9 Canada\\}

\date{\today}
\maketitle

\begin{abstract}

We study the 4-point function in the Keldysh formalism of the closed time path formulation of real time finite temperature field theory.  We derive the KMS conditions for these functions and discuss the number of 4-point functions that are independent.  We define a set of `physical' functions which are linear combinations of the usual Keldysh functions.  We show that these functions satisfy simple KMS conditions.  In addition, we consider a set of integral equations which represent a resummation of ladder graphs.  We show that these integral equations decouple when one uses the physical functions that we have defined. We discuss the generalization of these results to QED.  
\end{abstract}

\date{\today}

\vspace{3ex}

\section{Introduction}

Finite temperature field theory has been widely applied to 
 particle  physics, solid state physics and
the physics of the early universe. There are two formulations of finite temperature field theory: the  
imaginary time formulation (ITF) and the real time formulation (RTF)~\cite{kap,rep-145,Chou,Bellac}.
The imaginary time formalism involves the calculation of green functions with imaginary time arguments.  Physical green functions (for example, retarded and advanced green functions) have real time arguments.  In the ITF, at the end 
of the calculation, one must perform analytic continuations to obtain green functions with real time arguments.   For higher n-point functions, these analytic continuations become increasingly difficult.  

In contrast, in the RTF formalism  there is a simple and natural procedure for extracting the physical green functions.  In addition, the real time formalism can be generalized to non-equilibrium situations. However, the real time formalism involves a doubling of degrees of freedom. These  extra degrees of freedom become 
increasingly cumbersome to handle for higher n-point functions. 
Because of this doubling of degrees of freedom,  an n-point function in the RTF has $2^n$ components.  These components obey one 
constraint equation, which reduces the maximum number of independent components to $2^n-1$.  In equilibrium, the KMS conditions impose additional constraints \cite{Kubo,Martin}, reducing the maximum number of independent components to $2^{n-1} -1$. In addition, there are symmetries that arise from considering permutations of the external legs of an $n$-point function.  These symmetries exist only for legs that correspond to the same field, and consequently the number of constraints is different for different field theories.  

As an example, consider the 2-point function.  For $n=2$ we have $2^n-1=3$ components, which can be reduced to 2 using the well known KMS condition for the 2-point function \cite{peterH}.  Since, both legs of the 2-point function necessarily correspond to the same field, there is always an additional permutation symmetry  that reduces the number of independent functions to 1.  In the case of the 2-point function, the permutation symmetry is just the familiar relation between the retarded and advanced 2-point functions: $\Pi_R(P) = \Pi_A(-P)$.   The 3-point function has $2^n-1=7$ components which can be reduced to three independent functions using four KMS conditions.  The KMS conditions for the three-point functions  were derived in
the closed time path (CTP) representation from the spectral representations of the seven retarded-advanced functions \cite{megu}.  The number of additional constraints that result from the permutation symmetry will depend on what theory is being considered. 

In this paper we study the generalization of these rsults to the 4-point function.  The paper is organized as follows.  In section II we introduce the Keldysh definitions of the 4-point functions and derive the KMS conditions which relate these functions.  In section III we discuss the constraints that result from the permutation symmetry in a scalar theory.  In section IV we define the `physical' functions as linear combinations of the usual Keldysh functions.  We derive the KMS conditions for these `physical' functions and show that they are considerably simpler than the KMS conditions for the usual Keldysh functions. We also look at a set of integral equations that represent the resummation of ladder graphs and show that this set of equations decouples when the `physical' functions are used.   In section V we discuss the generalization of these results to QED, and in Section VI we present our conclusions.

\section{real time 4-point functions}


We start with a simple scalar theory.  The connected 4-point function is given by the contour ordered expectation value,
\bea
M^C_{abcd}(X,Y,Z,W) = \langle T_c \phi_a(X) \phi_b(Y) \phi_c(Z) \phi_d(W)\rangle
\eea
where the indices $\{a,b,c,d\}$ take values $\{1,2\}$ corresponding to the two branches of the CTP contour.  Due to the constraint 
\begin{equation}
\sum_{a,b,c,d=1}^{2} (-1)^{a+b+c+d} M^C_{abcd} = 0 \nonumber
\end{equation}
there are at most 15 independent components.  There are several different representations for these 15 components: 1) One can use the CTP functions themselves ($M_{1111}$, $M_{1112}$, $M_{1121}$, $\cdots$); 2) The $RA$ notation  uses real time green functions that are directly related to the green functions that one obtains from doing analytic continuations on the ITF green functions \cite{PA}; 3) The green functions can be written in the Keldysh representation using a tensor decomposition \cite{Keld,peterH}.  In this paper we will use the third option and write the 4-point function as the sum of 15 terms, each of which is a 16 component tensor which can be written as the
outer product of four 2-component vectors.  The 1PI four-point function is obtained by truncating external legs and can also be written as a sum of 16 component 
tensors.  We write,
\bea
8M &&= M_{R1}{1\choose -1}{1\choose 1}{1\choose 1}{1\choose 1} + 
M_{R2} {1\choose 1}{1\choose -1}{1\choose 1}{1\choose 1} + 
M_{R3} {1\choose 1}{1\choose 1}{1\choose -1}{1\choose 1} \nonumber \\
&&+ M_{R4} {1\choose 1}{1\choose 1}{1\choose 1}{1\choose -1} + 
M_A {1\choose 1}{1\choose 1}{1\choose -1}{1\choose -1} + 
M_B {1\choose -1}{1\choose 1}{1\choose -1}{1\choose 1} \nonumber\\
&&+ M_C {1\choose 1}{1\choose -1}{1\choose -1}{1\choose 1} + 
M_D {1\choose 1}{1\choose -1}{1\choose 1}{1\choose -1} + M_E {1\choose -1}{1\choose -1}{1\choose 1}{1\choose 1} \label{decompM}\\
&&+ M_F {1\choose -1}{1\choose 1}{1\choose 1}{1\choose -1} + 
M_{\alpha} {1\choose 1}{1\choose -1}{1\choose -1}{1\choose -1} + 
M_\beta {1\choose -1}{1\choose 1}{1\choose -1}{1\choose -1} \nonumber\\
&&+ M_\gamma {1\choose -1}{1\choose -1}{1\choose 1}{1\choose -1} + 
M_\delta {1\choose -1}{1\choose -1}{1\choose -1}{1\choose 1} + 
M_T {1\choose -1}{1\choose -1}{1\choose -1}{1\choose -1}\nonumber
\eea
There are several advantages to this choice of notation.  The 15 functions $M_{R1}, ~M_{R2}, ~M_{R3},$  $M_{R4},~M_A,~M_B,$ $M_C,~M_D,~M_E,$ $M_F,~M_\alpha,~M_\beta,$ $M_\gamma,~M_\delta,~M_T$ can be easily written in terms of the CTP functions.  It is straightforward to show that the first four of these functions $M_{Ri},~\{i=1,2,3,4\}$ are just the usual retarded 4-point functions. In addition, in a previous paper, we have developed a $Mathematica$ program for evaluating real time
Feynman amplitudes which uses this representation\cite{meghj}. This program substantially reduces the technical difficulties associated with the real time formalism, and makes it possible for us to exploit its advantages.

\subsection{The KMS condition}


In momentum space one can derive a set of relations for the connected functions directly from the cyclic property of the trace.  It is always true that,
\bea
\label{tilde1}
M^C_{1111}&=&M^{C*}_{2222}
\eea
and in equilibrium we have also,
\bea
\label{tilde2}
M^{C*}_{2111}&=&e^{\beta P^0_1}M^C_{1222}, \, \, \, M^{C*}_{1211}=e^{\beta P^0_2}M^C_{2122}\nonumber \\
M^{C*}_{1121}&=&e^{\beta P^0_3}M^C_{2212}, \, \, \, M^{C*}_{1112}=e^{\beta P^0_4}M^C_{2221}\nonumber \\
M^{C*}_{2211}&=&e^{\beta( P^0_1+P^0_2)}M^C_{1122}\\
M^{C*}_{2121}&=&e^{\beta( P^0_1+P^0_3)}M^C_{1212}\nonumber\\
M^{C*}_{2112}&=&e^{\beta( P^0_1+P^0_4)}M^C_{1221}\nonumber
\eea
These equations (\ref{tilde1}) and (\ref{tilde2}) give  eight constraint conditions relating the components  $M^C_{abcd}; \{a,b,c,d=1,2\}$.   The number of constraints relating the 1PI functions is the same.  It can be shown that if the 1PI vertex is defined appropriately (in this case without any extra factors of `$i$'), the form of the constraints for the 1PI functions is identical.  Thus the maximum number if independent components is reduced from 15 to 7.

We can rewrite the KMS conditions (\ref{tilde1}) and (\ref{tilde2}) in terms of the 15 functions $M_i$ defined in equation (\ref{decompM}). We use the notation,
\bea
\label{BE}
n=\frac{1}{e^{\beta p_0} -1}\,;~~~~N=1+2n\,.
\eea
We obtain,  
\bea
M_A&=&N^{34}_{12}(M_E^*+N_1M_{R2}^*+N_2M_{R1}^*) -N_3 M_{R4} - N_4 M_{R3}\nonumber \\
M_C&=&N^{23}_{14}(M_F^*+N_1M_{R4}^*+N_4M_{R1}^*) - N_2M_{R3}-N_3M_{R2}
\nonumber \\
M_D&=&N^{24}_{13}(M_B^*+N_1M_{R3}^*+N_3M_{R1}^*) -  N_2M_{R4}-N_4M_{R2}\nonumber \\
M_\alpha & =&-M_A N_2 - M_D N_3 - M_{R4} N_2 N_3 - M_C N_4 - M_{R3} N_2 N_4 - M_{R2} N_3 N_4 \nonumber \\
& & +  M_{R1}^* (1 + N_2 N_3 + N_2 N_4 + N_3 N_4) \nonumber \\
M_\beta  & =&-M_A N_1 - M_F N_3 - M_{R4} N_1 N_3 - M_B N_4 - M_{R3} N_1 N_4 - M_{R1} N_3 N_4  \nonumber \\ 
 & & + M_{R2}^* (1 + N_1 N_3 + N_1 N_4 + N_3 N_4) \nonumber \\
M\gamma & =&-M_D N_1 - M_F N_2 - M_{R4} N_1 N_2 - M_E N_4 - M_{R2} N_1 N_4 - M_{R1} N_2 N_4
\nonumber \\
 & &+ M_{R3}^* (1 + N_1 N_2 + N_1 N_4 + N_2 N_4)  \label{kmsjohn1} \\
M_\delta & =&-M_C N_1 - M_B N_2 - M_{R3} N_1 N_2 - M_E N_3 - M_{R2} N_1 N_3 - M_{R1} N_2 N_3  \nonumber \\
  & &+M_{R4}^* (1 + N_1 N_2 + N_1 N_3 + N_2 N_3) \nonumber \\
M_T &=& M_A N_1 N_2 + M_D N_1 N_3 + M_F N_2 N_3 + 2 M_{R4} N_1 N_2 N_3 + M_C N_1 N_4 
 \nonumber \\
& &+ M_B N_2 N_4 +  2 M_{R3} N_1 N_2 N_4 + M_E N_3 N_4  \nonumber \\
& & + 2 M_{R2} N_1 N_3 N_4 + 2 M_{R1} N_2 N_3 N_4   \nonumber \\
 & & - M_{R4}^* N_4 (1 + N_1 N_2 + N_1 N_3 + N_2 N_3)   \nonumber \\
  & & - M_{R3}^* N_3 (1 + N_1 N_2 + N_1 N_4 + N_2 N_4)  \nonumber \\
  & & - M_{R2}^* N_2 (1 + N_1 N_3 + N_1 N_4 + N_3 N_4)   \nonumber \\
  & & - M_{R1}^* N_1 (1 + N_2 N_3 + N_2 N_4 + N_3 N_4)  
\eea
where
\bea
N^{kl}_{ij} = \frac{N_k+N_l}{N_i+N_j} 
\eea

\section{Permutation Symmetry}

We have shown above that there are at most 7 independent 4-point functions.  In this section we will show that there is a permutation symmetry that further reduces the number of independent functions.  Ignoring the KMS conditions for the moment, we consider the 15 functions defined in equation (\ref{decompM}).  We note that the four $M_{R1},\cdots M_{R4}$ all have one minus sign among the lower components of the two component vectors involved in the decomposition.  Similarly, the six functons $M_A,\cdots M_F$ have two minus signs, the four $M_\alpha,\cdots M_\delta$ have three minus signs, and $M_T$ has four minus signs.  It is straightforward to show that vertices with the same number of minus signs, which we refer to as a given `type,' are related to each other through a permutation symmetry.  

We consider permutations of the external legs. Notice that there are many such permutations that don't produce any constraints.  For example, starting from the definition in co-ordinate space,
\bea
&&M_{R1}(X_1,X_2,X_3,X_4) \\
&&~~~~~= \sum_{\{2,3,4\}}\theta(X_1^0-X_2^0) \theta(X_2^0-X_3^0) \theta(X_3^0-X_4^0) \langle T_c [\phi(X_1),[\phi(X_2),[\phi(X_3),\phi(X_4)]]]\rangle
\nonumber 
\eea
where the sum is over all permutations of the coordinates $\{X_1,X_2,X_3\}$, 
it is easy to see that 
\bea
M_{R1}(P_1,P_2,P_3,P_4) = M_{R1}(P_1,P_2,P_4,P_3) = \cdots = M_{R1}(P_1,\{P_2,P_3,P_4\})
\eea
where the curly brackets indicate any permutation of the last three momentum variables.  Non-trivial constraints arise from the relations,
\bea
&&M_{R1}(P_1,P_2,P_3,P_4) = M_{R2}(P_2,P_1,P_4,P_3) =M_{R3}(P_3,P_4,P_1,P_2) =M_{R4}(P_4,P_3,P_2,P_1) \nonumber \\
&&M_A(P_1,P_2,P_3,P_4) = M_B(P_4,P_2,P_3,P_1) = M_E(P_3,P_4,P_1,P_2) = M_D(P_2,P_4,P_1,P_3) \nonumber\\
&&~~~~~~= M_C(P_1,P_3,P_4,P_2) = M_F(P_3,P_2,P_1,P_4)\label{scalarperms} \\
&&M_{\alpha}(P_1,P_2,P_3,P_4) = M_{\beta}(P_2,P_1,P_4,P_3) =M_{\gamma}(P_3,P_4,P_1,P_2) =M_{\delta}(P_4,P_3,P_2,P_1) \nonumber 
\eea
These relations show that the maximum number of independent functions is given by the number of `types' of 4-point functions, which is four.  We can now consider how much this number is reduced by the eight KMS conditions.  It is easy to see that the three KMS conditions for $M_\beta$, $M_\gamma$ and $M_\delta$ can be obtained from the KMS condition for $M_\alpha$ using the permutation symmetry.  Similarly, for the functions $M_A,\cdots M_F$, two of the three KMS conditions can be obtained from the other.  Thus, the number of independent functons is two, which we can take to be $M_{R1}$ and $M_A$.

\section{The Physical Functions}

\subsection{Definitions}

The KMS conditions in the form (\ref{kmsjohn1}) are extremely messy.  We can obtain very simple, elegant expressons by identifying what we will call the `physical' combinations.  We refer to these combinations as physical because they are the ones that result in decoupled integral equations for ladder resummed quantities in the pinch approximation \cite{visco}, and because they satisfy simple KMS conditions.  These points will be discussed below.  
We make the following definitions:
\bea
&&\bar M_A = M_A + N_3 M_{R4} + N_4 M_{R3} \nonumber \\
&&\bar M_B = M_B + N_1 M_{R3} + N_3 M_{R1} \nonumber \\
&&\bar M_C = M_C + N_2 M_{R3} + N_3 M_{R2} \nonumber \\
&&\bar M_D = M_D + N_2 M_{R4} + N_4 M_{R2} \nonumber \\
&&\bar M_E = M_E + N_1 M_{R2} + N_2 M_{R1} \nonumber \\
&&\bar M_F = M_F + N_1 M_{R4} + N_4 M_{R1} \label{tildeMA} 
\eea
All of these expressions can be obtained from any one of the others by using the permutation symmetry.  Similarly we define, 
\bea
&&\bar M^{(CD)}_\alpha= M_\alpha+N_3 M_D +N_4 M_C + N_3 N_4 M_{R2}\nonumber \\
&&\bar M^{(AD)}_\alpha= M_\alpha+N_2 M_A +N_3 M_D + N_2 N_3 M_{R4}\nonumber \\
&&\bar M^{(AC)}_\alpha= M_\alpha+N_4 M_C +N_2 M_A + N_2 N_4 M_{R3}\nonumber \\
&&\bar M^{(BF)}_\beta= M_\beta+N_3 M_F +N_4 M_B + N_3N_4M_{R1}\nonumber \\
&&\bar M^{(AF)}_\beta= M_\beta+N_1 M_A +N_3 M_F + N_1N_3M_{R4}\nonumber \\
&&\bar M^{(AB)}_\beta= M_\beta+N_1 M_A +N_4 M_B + N_1N_4M_{R3}\nonumber \\
&&\bar M^{(DF)}_\gamma= M_\gamma+N_1 M_D +N_2 M_F + N_1N_2 M_{R4}\nonumber \\
&&\bar M^{(EF)}_\gamma= M_\gamma+N_4 M_E +N_2 M_F + N_2N_4 M_{R1}\nonumber \\
&&\bar M^{(DE)}_\gamma= M_\gamma+N_1 M_D +N_4 M_E + N_1N_4 M_{R2}\nonumber \\
&&\bar M^{(BC)}_\delta= M_\delta+N_1 M_C +N_2 M_B + N_1N_2 M_{R3}\nonumber \\
&&\bar M^{(CE)}_\delta= M_\delta+N_1 M_C +N_3 M_E + N_1N_3 M_{R2}\nonumber \\
&&\bar M^{(BE)}_\delta= M_\delta+N_2 M_B +N_3 M_E + N_2N_3 M_{R1}\label{tildeMalpha} 
\eea
Note that there are three possible definitions for each of the $\bar M_{\alpha , \cdots \delta}$, all of which are related to each other through the permutation symmetry.  Finally, we have six possible definitions of $\bar M_T$, which are related to each other through the permutation symmetry: 
\bea
&&\bar M^{(\alpha\beta)}_T  = M_T+N_1 M_\alpha +N_2 M_\beta + N_1N_2M_A\nonumber\\
&&\bar M^{(\alpha\gamma)}_T  = M_T+N_1 M_\alpha +N_3 M_\gamma +N_1N_3M_D\nonumber\\
&&\bar M^{(\alpha\delta)}_T  = M_T+N_1 M_\alpha +N_4 M_\delta + N_1N_4M_C\nonumber\\
&&\bar M^{(\beta\gamma)}_T  = M_T+N_2 M_\beta +N_3 M_\gamma + N_2N_3M_F\nonumber\\
&&\bar M^{(\beta\delta)}_T  = M_T+N_2 M_\beta +N_4 M_\delta + N_2N_4M_B\nonumber\\
&&\bar M^{(\gamma\delta)}_T   = M_T+N_3 M_\gamma +N_4 M_\delta + N_3N_4M_E\label{tildeMT}
\eea

\subsection{KMS conditions}

Using the `physical' definitions (\ref{tildeMA}), (\ref{tildeMalpha}) and (\ref{tildeMT}) we can rewrite the KMS conditions (\ref{tilde1}) and (\ref{tilde2}) or (\ref{kmsjohn1}) to obtain simple expressions.  In this section we give the results for one choice of definition for each of $\bar M_{\alpha, \cdots \delta,T}$.  The full set of results can be obtained using the permutation symmetry and is given in Appendix A.  
\bea
&&(N_1+N_2)\bar M_A =(N_3+N_4) \bar M_E^* \nonumber \\
&&(N_1+N_4)\bar M_C =(N_2+N_3) \bar M_F^* \label{KMStilde1} \\
&&(N_1+N_3) \bar M_D =(N_2+N_4) \bar M_B^* \nonumber \\
&&\bar M^{(CD)}_\alpha = -N_2 \bar M_A + M_{R1}^*(1+N_2N_3 + N_2 N_4 + N_3N_4)\nonumber \\
&&\bar M^{(AB)}_\beta = -N_3 \bar M_F + M_{R2}^*(1+N_1N_3 + N_1 N_4 + N_3N_4)\nonumber \\
&&\bar M^{(DF)}_\gamma = -N_4 \bar M_E + M_{R3}^*(1+N_1N_2 + N_1 N_4 + N_2N_4)\nonumber \\
&&\bar M^{(BC)}_\delta = -N_3 \bar M_E + M_{R4}^*(1+N_1N_2 + N_1 N_3 + N_2N_3)\nonumber \\
&&\bar M^{(\alpha\beta)}_T = N_3 N_4 \bar M_E - M^*_{R3}N_3 (1+N_1N_2 + N_1 N_4 + N_2N_4) \nonumber \\
&&~~~~~~~~- M_{R4}^* N_4 (1+N_1N_2 + N_1 N_3 + N_2N_3) \nonumber 
\eea

The usefulness of these expressions can be verified in a straightforward way by looking at the diagram shown in Fig. [1].  We can use the KMS conditions to verify the KMS relation $\Pi_F(Q) = N_Q(\Pi_R(Q) - \Pi_A(Q))$ for this diagram.    Using the $Mathematica$ program developed in \cite{meghj} we find,
\bea
\Pi_R(Q) = -\frac{\lambda^2}{4}\int \frac{d^4p}{(2\pi)^4} \frac{d^4k}{(2\pi)^4}\{&& M_{R1}(a_p f_k f_{p+k+q} + f_p a_k f_{p+k+q} + a_p a_k r_{p+k+q} + f_p f_k r_{p+k+q}) \nonumber \\
+ && M_B (a_p r_k f_{p+k+q} + f_p r_k r_{p+k+q})
+ M_E(r_p a_k f_{p+k+q} + r_p f_k r_{p+k+q}) \nonumber \\
+ && M_F ( a_p f_k a_{p+k+q} + f_p a_k a_{p+k+q})
+ M_\beta (a_p r_k a_{p+k+q}) \nonumber \\
 + && M_\gamma( r_p a_k a_{p+k+q}) 
+  M_\delta(r_p r_k r_{p+k+q}) \} \nonumber \\
\Pi_A(Q) = -\frac{\lambda^2}{4}\int \frac{d^4p}{(2\pi)^4} \frac{d^4k}{(2\pi)^4} \{ && M_{R2}(r_p f_k f_{p+k+q} + r_p a_k r_{p+k+q}) + M_{R3}(f_p r_k f_{p+k+q} \nonumber \\
+&& a_p r_k r_{p+k+q}) + M_{R4} ( f_p f_k a_{p+k+q} + a_p a_k a_{p+k+q}) + M_A(f_p r_k a_{p+k+q}) \nonumber \\ 
+ && M_C (r_p r_k f_{p+k+q}) 
+  M_D(r_p f_k a_{p+k+q}) + M_\alpha (r_p r_k a_{p+k+q}) \} \nonumber \\
\Pi_F(Q) = -\frac{\lambda^2}{4}\int \frac{d^4p}{(2\pi)^4} \frac{d^4k}{(2\pi)^4}\{ &&M_{R1}(a_p a_k f_{p+k+q} + f_p f_k f_{p+k+q} + a_p f_k r_{p+k+q} + f_p a_k r_{p+k+q}) \nonumber \\
+ && M_F(a_p a_k a_{p+k+q} + f_p f_k a_{p+k+q})
+M_B(f_p r_k f_{p+k+q} + a_p r_k r_{p+k+q}) \nonumber \\
+ && M_E(r_p f_k f_{p+k+q} + r_p a_k r_{p+k+q})
+ M_\beta (f_p r_k a_{p+k+q}) \nonumber \\
+ && M_\gamma(r_p f_k a_{p+k+q}) 
+  M_\delta(r_p r_k f_{p+k+q})
+M_T(r_p r_k a_{p+k+q}) \}
\eea
where the vertices inside the integrals are defined as $M:= M(Q,P,K,-P-K-Q)$ and we have  used the short hand notation for the propagators $D_R(P) = r_p$ etc. For simplicity, we will look at just one term.  We choose the term proportional to $r_p r_k a_{p+k+q}$.  It is easy to show that we have
\bea
&& \Pi_R(Q) = 0 \cdot r_p r_k a_{p+k+q} + \cdots \nonumber \\
&& \Pi_A(Q) = -\frac{\lambda^2}{4}\int \frac{d^4p}{(2\pi)^4} \frac{d^4k}{(2\pi)^4}(N_3 \bar M_D + \bar M_\alpha^{(AC)}) r_p r_k a_{p+k+q}+ \cdots \nonumber \\
&& \Pi_F(Q) = -\frac{\lambda^2}{4}\int \frac{d^4p}{(2\pi)^4} \frac{d^4k}{(2\pi)^4}(N_3 \bar M_\gamma^{(EF)} + \bar M_T^{(\beta\delta)})r_p r_k a_{p+k+q} + \cdots 
\eea
Notice that these expressions are particularly simple using the physical vertices.  Using the KMS conditions given in Appendix A with $P_1=Q$, $P_2=P$, $P_3=K$ and $P_4= -(P+K+Q)$, it is easy to show that $\Pi_F(Q) = N_Q(\Pi_R(Q) - \Pi_A(Q))$ is satisfied for this term.  All of the other terms work exactly the same way.  It is clear that this calculation would be incredibly tedious wihout making use of the `physical' definitions.

\subsection{Decoupled integral equations}
The vertices defined in (\ref{tildeMA}), (\ref{tildeMalpha}) and (\ref{tildeMT}) result in decoupled integral equations for the ladder resummed vertices, in the pinch approximation.  Fig. [2] shows the integral equation whose solution gives the resummation of ladder contributions to the 4-point vertex.  We obtain the pinch approximation by taking the limit that $Q$ is small.  This limit is of physical interest in calculations of transport coefficients like the viscosity, where $Q$ is taken to zero at the end of the calculation \cite{visco}.  This limit gives rise to what is known as the `pinch effect.'  Terms with a product of factors $a_p a_{q-p}$  and $r_p r_{q-p}$ contain an extra factor $\sim 1/Q$ relative to terms with products of propagators $a_p r_{q-p}$ or $r_p a_{q-p}$.  The large terms occur when the integration contour is `pinched' between the poles of the two propagators, which gives rise to a factor in the denominator that is proportional to the imaginary part of these propagators.  Thus terms proportional to $a_p r_{q-p}$ or $r_p a_{q-p}$ can be dropped in the pinch approximation.  

To simplify notation, we make the definitions
\bea
&&\delta = r_{p'} a_r + a_{p'} r_r + f_{p'} f_r \nonumber \\
&&\phi_1 = a_{p'} f_r + f_{p'} r_r  \\
&&\phi_2 = r_{p'} f_r + f_{p'} a_r \nonumber
\eea
where $P' = R+Q-P_1-P$.  
The integral equation for $M_T$ has the form,
\bea
&&M_T(P_1,Q-P_1,P_3,P_4) = -\frac{\lambda^2}{4}\int \frac{d^4p}{(2\pi)^4} \frac{d^4r}{(2\pi)^4} \nonumber \\
&&~~[a_{q-p}a_p M_A(P,Q-P,P3,P4) (\delta N_{q-p} N_p - N_{q-p}\phi_1 - N_p \phi_2) \nonumber \\
&&~~+ r_{q-p} r_p ~\delta ~(M_A(P,Q-P,P3,P4)  N_{q-p}N_p + M_\alpha(P,Q-P,P3,P4)  N_p \nonumber \\
&&~~+ M_\beta(P,Q-P,P3,P4)  N_{q-p} + M_T(P,Q-P,P3,P4) )] 
\eea
  We can compare this result with that obtained for the `physical' vertex $\bar M_T$.  We have,
\bea
\bar M^{(\alpha\beta)}_T(P_1,Q-P_1,P_3,P_4) =&&-\frac{\lambda^2}{4}\int \frac{d^4p}{(2\pi)^4} \frac{d^4r}{(2\pi)^4} 
r_p r_{q-p}\nonumber \\
&&[\delta + N_1 \phi_2 + N_2 \phi_1] \bar M^{(\alpha\beta)}_T(P,Q-P,P3,P4) 
\eea
Thus the `physical' vertex $\bar M_T$ satisfies a simple decoupled integral equation. The other `physical' vertices defined in (\ref{tildeMA}) and (\ref{tildeMalpha}) also satisfy decoupled integral equations.

\section{QED}

In this section we discuss the generalization of these results to QED.  In QED there are two kinds of fields, fermion fields and photon fields.  This multiplicity of fields introduces three kinds of complications: 1) there are three different 4-point functions; 2) the thermal distributions for fermion fields are different from the distributions for boson fields (like the scalar field discussed in the first part of this paper); 3) the allowed permutations are restricted to the interchange of fields of the same kind, which reduces the number of symmetry relations. We discuss these differences below.  

There are  
three different 4-point functions: the 4-photon vertex, the 4-fermion vertex and the 2-photon / 2-fermion vertex.  These four vertices are written,
\bea
M^{C\mu\nu\lambda\tau}_{abcd}(X,Y,Z,W) &=& \langle T_c A^\mu_a(X) A^\nu_b(Y) A^\lambda_c(Z) A^\tau_d(W) \rangle \nonumber \\
M^{C\mu\nu}_{abcd}(X,Y,Z,W) &=& \langle T_c \psi_a(X) \psi_b^\dagger(Y) A^\mu_c(Z) A^\nu_d(W) \rangle  \\
M^{C}_{abcd}(X,Y,Z,W) &=& \langle T_c \psi_a(X) \psi_b(Y) \psi^\dagger _c(Z) \psi^\dagger_d(W) \rangle \nonumber 
\eea
In momentum space, the variables $\{P_1,P_2,P_3,P_4\}$ correspond to the coordinate variables $\{X,Y,Z,W\}$, respectively. The KMS conditions are the same as (\ref{kmsjohn1}) except that the thermal factors which contain Bose-Einstein distributions (\ref{BE}) are replaced by factors that contain Fermi-Dirac distributions, when the corresponding momentum is carried by a fermion.  We define,
\bea
\label{FD}
n_F = \frac{1}{e^{\beta p_0}+1}\,;~~~~N_F =1 - 2n_F\,. 
\eea

For a theory that involves more than one kind of field, the symmetry relations (\ref{scalarperms}) are more complicated. We will consider each vertex in turn.  

\noindent {\bf 1.)} We start with the 2-photon / 2-fermion vertex. It is straightforward to show that these vertices obey the relations,
\bea
&&M^{\mu\nu}_{R1}(P_1,P_2,P_3,P_4) = M^{\dagger\mu\nu}_{R2}(P_2,P_1,P_3,P_4)  \nonumber \\
&&M^{\mu\nu}_{R3}(P_1,P_2,P_3,P_4) = M^{\nu\mu}_{R4}(P_1,P_2,P_4,P_3) \nonumber \\
&&M^{\mu\nu}_{C}(P_1,P_2,P_3,P_4) = M^{\nu\mu}_D(P_1,P_2,P_4,P_3) = M^{\dagger\mu\nu}_B(P_2,P_1,P_3,P_4) = M^{\dagger\nu\mu}_F(P_2,P_1,P_4,P_3) \nonumber \\
&&M^{\mu\nu}_{\alpha}(P_1,P_2,P_3,P_4) = M^{\dagger\mu\nu}_{\beta}(P_2,P_1,P_3,P_4)  \nonumber \\
&&M^{\mu\nu}_{\gamma}(P_1,P_2,P_3,P_4) = M^{\nu\mu}_{\delta}(P_1,P_2,P_4,P_3) \label{2p2fperms}
\eea
Thus, the maximum number of independent components is 8.  For example, we can take the independent 4-point functions to be, $M^{\mu\nu}_{R1},~M^{\mu\nu}_{R3},~M^{\mu\nu}_{\alpha},~M^{\mu\nu}_\gamma,~M^{\mu\nu}_A,~M^{\mu\nu}_E,~M^{\mu\nu}_B,~M^{\mu\nu}_T$.  If the KMS conditions are used, it is straightforward to show that $M^{\mu\nu}_E$ can be obtained from $M^{\mu\nu}_A$, $M^{\mu\nu}_{R1}$ and $M^{\mu\nu}_{R3}$, and that $M^{\mu\nu}_{\alpha},~M^{\mu\nu}_\gamma,$ and $M^{\mu\nu}_T$ can be obtained from $M^{\mu\nu}_A$, $M^{\mu\nu}_B$, $M^{\mu\nu}_{R1}$ and $M^{\mu\nu}_{R3}$.  Thus, the number is independent functions is four, which can be taken to be $M^{\mu\nu}_{R1}$, $M^{\mu\nu}_{R3}$, $M^{\mu\nu}_A$, and $M^{\mu\nu}_B$.

\noindent {\bf 2.)} Next we consider the 4-photon vertex. The permutation relations are,
\bea
&&M_{R1}^{\mu\nu\lambda\tau}(P_1,P_2,P_3,P_4) = M_{R2}^{\nu\mu\lambda\tau}(P_2,P_1,P_3,P_4) =M_{R3}^{\lambda\tau\mu\nu}(P_3,P_4,P_1,P_2) =M_{R4}^{\tau\lambda\nu\mu}(P_4,P_3,P_2,P_1) \nonumber \\
&&M_A^{\mu\nu\lambda\tau}(P_1,P_2,P_3,P_4) = M_B^{\tau \nu \lambda \mu}(P_4,P_2,P_3,P_1) = M_E^{\lambda\tau \mu\nu}(P_3,P_4,P_1,P_2) = M_D^{\nu \tau \mu\lambda}(P_2,P_4,P_1,P_3) \nonumber\\
&&~~~~~~= M_C^{\mu\lambda\tau\nu}(P_1,P_3,P_4,P_2) = M_F^{\lambda\nu\mu\tau}(P_3,P_2,P_1,P_4)\label{4photonperms} \\
&&M_{\alpha}^{\mu\nu\lambda\tau}(P_1,P_2,P_3,P_4) = M_{\beta}^{\nu \mu\tau\lambda}(P_2,P_1,P_4,P_3) =M_{\gamma}^{\lambda\tau\mu\nu}(P_3,P_4,P_1,P_2) =M_{\delta}^{\tau\lambda\nu \mu}(P_4,P_3,P_2,P_1) \nonumber 
\eea
The maximum number of independent functions is four, which can be taken to be $M^{\mu\nu\lambda\tau}_{R1},$ $M^{\mu\nu\lambda\tau}_{A},$ $M^{\mu\nu\lambda\tau}_{\alpha}$ and $M^{\mu\nu\lambda\tau}_{T}$. Using the KMS conditions, $M^{\mu\nu\lambda\tau}_{\alpha}$ and $M^{\mu\nu\lambda\tau}_{T}$ are not independent, which reduces the number of independent functions to two.

\noindent {\bf 3.)} Finally we consider the 4-fermion vertex. The permutation relations are,
\bea
&&M_{R1}(P_1,P_2,P_3,P_4) = M_{R2}(P_2,P_1,P_3,P_4) = M^\dagger_{R3}(P_3,P_4,P_1,P_2) = M_{R4}^\dagger(P_4,P_3,P_2,P_1) \nonumber \\
&&M_A(P_1,P_2,P_3,P_4) = M_E^\dagger(P_3,P_4,P_1,P_2) \nonumber \\
&&M_C(P_1,P_2,P_3,P_4) = M_D(P_1,P_2,P_4,P_3) = M_B(P_2,P_1,P_3,P_4) = M_F(P_2,P_1,P_4,P_3) \nonumber \\
&&M_{\alpha}(P_1,P_2,P_3,P_4) = M_{\beta}(P_2,P_1,P_3,P_4) = M^\dagger_\gamma(P_3,P_4,P_1,P_2) = M^\dagger_\delta(P_4,P_3,P_2,P_1) \label{4fermionperms}
\eea
The maximum number of independent vertices is five, which can be taken as $M_{R1}$, $M_A$, $M_C$, $M_\alpha$ and $M_T$.  Using the KMS conditions, $M_\alpha$ and $M_T$ are not independent, which reduces the number of independent functions to three.

\section{Conclusion}

The real time formalism of finite temperature field theory has recently gained in popularity because it avoids the need
for analytical continuations that plagues the imaginary formalism. However, because of the extra degrees of freedom introduced in the RTF, calculations can be extremely tedious. In the Keldysh representation the 4-point function is written in terms of 15 separate components.  However, in equilibrium, the situation is in principle simplified by the existence of the KMS conditions.  For any n-point function, which can be expressed as a thermal expectation value of a set of operators, these conditions can be derived  using the cyclic property of the trace.  The KMS conditions have been derived previously in the Keldysh representation for the 2-point and 3-point functions \cite{peterH,megu}.  In this paper we have derived a set of KMS conditions for the Keldysh 4-point functions.  As is typical in the Keldysh representation, these equations are extremely complicated.  However, we have been able to identify linear combinations of the Keldysh functions which we call `physical' functions.  We have shown that the KMS conditions satisfied by these physical functions are relatively simple, and we have done two examples which demonstrate the usefulness of these `physical' definitions for doing calculations in the Keldysh representation. 

\vspace*{1cm}

{\bf Acknowledgments}

\noindent The authors gratefully acknowledge useful discussions with U. Heinz and R. Kobes.

\newpage
{\centerline {\large {\bf Appendix A}}}

\vspace*{.5cm}

\bea
&&(N_1+N_2)\bar M_A =(N_3+N_4) \bar M_E^* \nonumber \\
&&(N_1+N_4)\bar M_C =(N_2+N_3) \bar M_F^* \nonumber \\
&&(N_1+N_3) \bar M_D =(N_2+N_4) \bar M_B^* \nonumber \\
&&\bar M^{(CD)}_\alpha = -N_2 \bar M_A + M_{R1}^*(1+N_2N_3 + N_2 N_4 + N_3N_4)\nonumber \\
&&\bar M^{(AC)}_\alpha = -N_3 \bar M_D + M_{R1}^*(1+N_2N_3 + N_2 N_4 + N_3N_4)\nonumber \\
&&\bar M^{(AD)}_\alpha = -N_2 \bar M_C + M_{R1}^*(1+N_2N_3 + N_2 N_4 + N_3N_4)\nonumber \\
&&\bar M^{(AB)}_\beta = -N_3 \bar M_F + M_{R2}^*(1+N_1N_3 + N_1 N_4 + N_3N_4)\nonumber \\
&&\bar M^{(AF)}_\beta = -N_4 \bar M_B + M_{R2}^*(1+N_1N_3 + N_1 N_4 + N_3N_4)\nonumber \\
&&\bar M^{(BF)}_\beta = -N_1 \bar M_A + M_{R2}^*(1+N_1N_3 + N_1 N_4 + N_3N_4)\nonumber \\
&&\bar M^{(DF)}_\gamma = -N_4 \bar M_E + M_{R3}^*(1+N_1N_2 + N_1 N_4 + N_2N_4)\nonumber \\
&&\bar M^{(EF)}_\gamma = -N_1 \bar M_D + M_{R3}^*(1+N_1N_2 + N_1 N_4 + N_2N_4)\nonumber \\
&&\bar M^{(DE)}_\gamma = -N_2 \bar M_F + M_{R3}^*(1+N_1N_2 + N_1 N_4 + N_2N_4)\nonumber \\
&&\bar M^{(BC)}_\delta = -N_3 \bar M_E + M_{R4}^*(1+N_1N_2 + N_1 N_3 + N_2N_3)\nonumber \\
&&\bar M^{(CE)}_\delta = -N_2 \bar M_B + M_{R4}^*(1+N_1N_2 + N_1 N_3 + N_2N_3)\nonumber\\
&&\bar M^{(BE)}_\delta = -N_1 \bar M_C + M_{R4}^*(1+N_1N_2 + N_1 N_3 + N_2N_3) \nonumber \\
&&\bar M^{(\alpha\beta)}_T = N_3 N_4 \bar M_E - M^*_{R3}N_3 (1+N_1N_2 + N_1 N_4 + N_2N_4) - M_{R4}^* N_4 (1+N_1N_2 + N_1 N_3 + N_2N_3) \nonumber \\
&&\bar M^{(\alpha\gamma)}_T = N_2 N_4 \bar M_B - M^*_{R2}N_2 (1+N_1N_3 + N_1 N_4 + N_3N_4)  - M_{R4}^* N_4 (1+N_1N_2 + N_1 N_3 + N_2N_3) \nonumber \\
&&\bar M^{(\alpha\delta)}_T = N_2 N_3 \bar M_F - M^*_{R2}N_2 (1+N_1N_3 + N_1 N_4 + N_3N_4) - M_{R3}^* N_3 (1+N_1N_2 + N_1 N_4 + N_2N_4) \nonumber \\
&&\bar M^{(\beta\gamma)}_T = N_1 N_4 \bar M_C - M^*_{R1}N_1 (1+N_2N_3 + N_2 N_4 + N_3N_4)  - M_{R4}^* N_4 (1+N_1N_2 + N_1 N_3 + N_2N_3) \nonumber \\
&&\bar M^{(\beta\delta)}_T = N_1 N_3 \bar M_D  - M^*_{R1} N_1 (1+N_2N_3 + N_2 N_4 + N_3N_4) + M^*_{R3}N_3 (1+N_1N_2 + N_1 N_4 + N_2N_4)  \nonumber \\
&&\bar M^{(\gamma\delta)}_T = N_1 N_2 \bar M_A - M^*_{R1}N_1 (1+N_2N_3 + N_2 N_4 + N_3N_4) - M_{R2}^* N_2 (1+N_1N_3 + N_1 N_4 + N_3N_4) \nonumber  
\eea

\newpage

 \begin{figure}
\epsfxsize=4cm
\centerline{\epsfbox{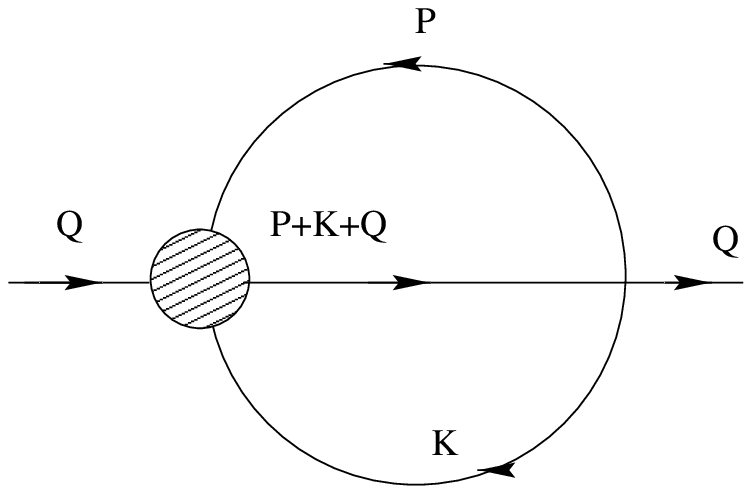}}
\vskip 0.4cm
 \caption{Sunset self-energy with corrected four-point vertex}
 \label{F1}
 \end{figure}

\begin{figure}
\epsfxsize=6cm
\centerline{\epsfbox{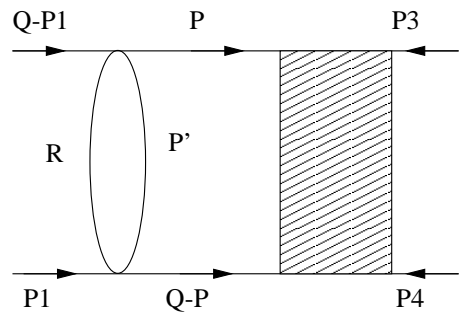}}
\vskip 0.4cm
 \caption{ladder resummation with corrected four-point vertex}
 \label{F2}
 \end{figure}

\end{document}